\documentclass[12pt,titlepage]{article}
\usepackage{amsmath}
\usepackage{verbatim}                            
\usepackage[super,sort&compress]{natbib}
\usepackage{hyperref}

\usepackage{graphicx}                            
\usepackage{ctable}                              
\usepackage[margin=1in]{geometry}                
\usepackage{pslatex}                             
\usepackage{setspace}                            
\usepackage{lmodern}
\usepackage{xcolor}
\title{An improved aerodynamic model for quasi-steady simulations of animal flight at moderate Reynolds numbers}
\author{Yunxing Su\footnotemark[1]\footnotemark[2] \and Xiaozhou Fan\footnotemark[1]\footnotemark[3] \and Kyohei Onoue\footnotemark[4]\and Hamid Vejdani\footnotemark[5]\and  Kenneth S. Breuer\footnotemark[6]}
\footnotetext[1]{Yunxing Su and Xiaozhou Fan contributed equally to the present work} 
\footnotetext[2]{Current address: Research associate at the University of Colorado Boulder}
\footnotetext[3]{Author for correspondence (email: xzfan@caltech.edu). Current address: postdoc at Caltech}
\footnotetext[4]{Senior aerodynamics lead engineer, Honda aircraft company}
\footnotetext[5]{Associate Professor, mechanical engineering department at Lawrence Technological University}
\footnotetext[6]{Professor of Engineering and Professor of Ecology, Evolution, and Organismal Biology, and Director of the Center of Fluid Mechanics at Brown University}
\date{}

\begin{document}
\maketitle

\section*{Abstract}
We report on experimental and numerical studies aimed at developing an improved paradigm to model animal flight at moderate Reynolds numbers ($ 20 k - 50k $). A series of experiments were performed to characterize the behaviors of aerodynamic forces and moment associated with a quasi-steady rectangular wing over a range of angle of attack, $\alpha$. We demonstrate that, while the drag coefficient curve, $C_D(\alpha)$, can be accurately modeled solely by a simple trigonometric function, the evolution of lift coefficient curve, $C_L(\alpha)$, is governed by the sum of trigonometric and exponential functions, where the latter captures the linear variation in lift coefficient within the small-angle regime, as predicted by the linear inviscid theory. In addition, we establish an empirical relation between the location of the center of pressure and $\alpha$, which can be used in conjunction with the proposed aerodynamic formulas (i.e., $C_L$ and $C_D$) to evaluate the pitching moment coefficient, $C_M(\alpha)$, about any arbitrary axis. These quasi-steady formulations are then utilized within a previously tested flapping-wing code to simulate the forward flight of a pigeon and a bat at various flight speeds, and the results are compared against previously reported experimental data. We successfully demonstrate that the proposed formulas yield much better agreement with wingbeat frequency for both animals, especially at higher flight speeds. In addition, the small-angle regime proves critical in offering higher $C_L/C_D$, leading to solutions with lower power consumption and body pitching variation, both of which are important aspects in designing future flapping wing robots.

\section*{Key words}
Reduced-order modeling; Vortex dynamics; Animal locomotion; Experimental fluid mechanics

\section*{Introduction}
There has been a substantial interest in understanding the physics behind birds and bats flapping flight, both to understand the intricate mechanics of these natural fliers and to pave the way for the development of Micro Air Vehicles (MAVs) that mirror their counterparts in offering superior maneuverability, efficiency and payload capabilities \citep{shyy2013,Shyy2016}. Significant advancement has been made in understanding the aerodynamics of flapping flight \citep{mueller2001fixed,wang2005dissecting,ansari2006aerodynamic,shyy2008aerodynamics,taha2012flight,Fan2024a,Fan2025}. The existing research is predominantly focused on relatively small fliers such as insects, characterized by Reynolds numbers less than $1000$ \citep[e.g.,][]{Dickinson1993,Dickinson1999,Sane2001,Sane2002,berman2007energy}. However, when examining larger fliers like birds and bats, the Reynolds number substantially increases to approximately 40,000 \citep{hedrick2006flight,Vejdani2019,fan2021a,fan2022}. Given that a significant proportion of contemporary MAVs operate within this intermediate range of Reynolds numbers, a computationally-efficient aerodynamic model that connects the wing aerodynamic forces to the angle of attack, $\alpha$, is required for designing controllers and studying the stability for these vehicles.

Various aerodynamic models with distinct characteristics are employed in the examination of animal flight. For example, element-wise aerodynamic models proposed by \citet{Dickinson1999,Sane2001,Sane2002} have been extensively used for modeling aerodynamic forces \citep[e.g., see,][]{Parslew2010, Parslew2015, Cheng2016, Diana2016} due to their simple nature and low computational cost, making it highly practical for tasks such as dynamical modeling, control and optimization calculations. \citet{Cheng2016} employ this quasi-steady modeling together with blade-element theory to analyze the flight mechanics and control of escaping maneuvers in hummingbirds. The authors applied a framework for translational lift and drag similar to \citet{Dickinson1999}, but adjusted slightly the numerical values of the aerodynamic model specifically for hummingbirds with  insights from CFD simulations \citep{song2014comparison}. 
\citet{berman2007energy} used blade-element theory together with the quasi-steady model \citep{Dickinson1999,Sane2002} to calculate the aerodynamic forces for their modeling of hawkmoth and fruitfly. Using this model, the authors \citep{berman2007energy} successfully identified the most efficient wing movements of hovering insect flight that minimize power consumption while ensuring sufficient lift to maintain a consistent average altitude throughout a single flapping period.

However, the simple sinusoidal aerodynamic force coefficients used in  this model \citep{Dickinson1999} fall short of fully encapsulating the genuine aerodynamic performance at the moderate to high Reynolds number range of $10^4 - 10^5$, typical for larger birds or bats at cruise speed. \citet{ellington2006insects} reported that the high force coefficients observed in the flapping wings of insects \citep{ellington1996leading,Dickinson1999,usherwood2002aerodynamicsa,usherwood2002aerodynamicsb} may not persist at Reynolds numbers relevant for bird flight (hummingbird may be an aerodynamic transitional regime between insects and birds). Consequently, numerous researchers have customized and refined this model to better align with their specific requirements and enhance its accuracy. In particular,  \citet{Parslew2015} performed panel method computations to capture the high-lift-to-drag ratios experienced at low angles of attack ($-18^{\circ}$ to $14^{\circ}$) and successfully captured the pigeon cruise flight mode. Based on their aerodynamic model, \citet{Parslew2010} and \citet{Parslew2015} successfully developed a predictive model to analyze pigeon flight for different flight modes (cruise, climbing, and descending) over a range of forward speeds. 

The Dickinson and Sane model was developed based on the experiments mimicking the aerodynamic performance of insect flight at a Reynolds number of the order of 100. However, \citet{Dickinson1993} shows that the aerodynamic performance of translating wings is highly sensitive to Reynolds number.  At high Reynolds number, linear inviscid theory predicts that the slope of $C_L-\alpha$ should be $2\pi$ for small angle of $\alpha$ below the stall angle \cite{Anderson2000}.  In contrast,
\citet{Dickinson1999} measured a much lower value around $3.8$. 
When used at larger Reynolds number, the drag model $C_D - \alpha$ measured at low Reynolds number, also tends to be larger than true values, which may explain the lower-than-observed thrust predicted for cruising bat flight\citep{fan2022}. Similar direct applications of the low Re formulation to high Reynolds number cases may also be found in literature \citep{Byl2010, Chen2010, Parslew2010, Parslew2015, Stowers2015, Cheng2016, Read2016}.

Of course, researchers have also utilized approaches different from quasi-steady modeling to understand $C_L, C_D - \alpha$ at $Re \sim 10^4 - 10^5$. \citet{carruthers2010mechanics,carruthers2010aerodynamics} utilized an aerodynamic force model by performing panel method computations (XFOIL) with wing planforms reconstructed from images of flapping bird wings. \citet{usherwood2009aerodynamic} investigated the aerodynamic performance of a revolving, dried pigeon wing and a flat plate replica using both direct force transducer and pressure sensor measurements at Reynolds number up to 10,800. For both cases (dried pigeon wing and flat plate), high lift coefficients are obtained with the maximum lift force achieved at an angle of attack up to $43^{\circ}$. In addition, \citet{taha2014state} utilized a state-space representation of the unsteady aerodynamics of flapping flight and proposed a reduced-order model for flight dynamics and control. The model extends Duhamel's principle to unconventional lift curves, aiming specifically to incorporate the contribution of the Leading-Edge Vortex (LEV).

Nevertheless, quasi steady modelling retains considerable appeal for its simplicity and accuracy, and thus, in this study, we propose improvements to the Sane-Dickinson quasi-steady formulation for lift and drag that are relevant at the moderate to large Reynolds numbers pertinent to avian and bat flight. In addition, it has been proven challenging to accurately predict the pitching motion of a flapping wing flight robot or animals \citep{Windes2018,fan2021a} due to the lack of reliable data on the change of wing center of pressure with $\alpha$ (see exception for low Reynolds number \citep{han2015}).  To address this concern we also present pitching moment measurements over this intermediate Reynolds number range.

\section*{Materials and methods}

\subsection*{Experimental setup}

\begin{figure}[!t]    
\centerline{ \includegraphics[trim=0 0 0 0,clip,width=5.5in]{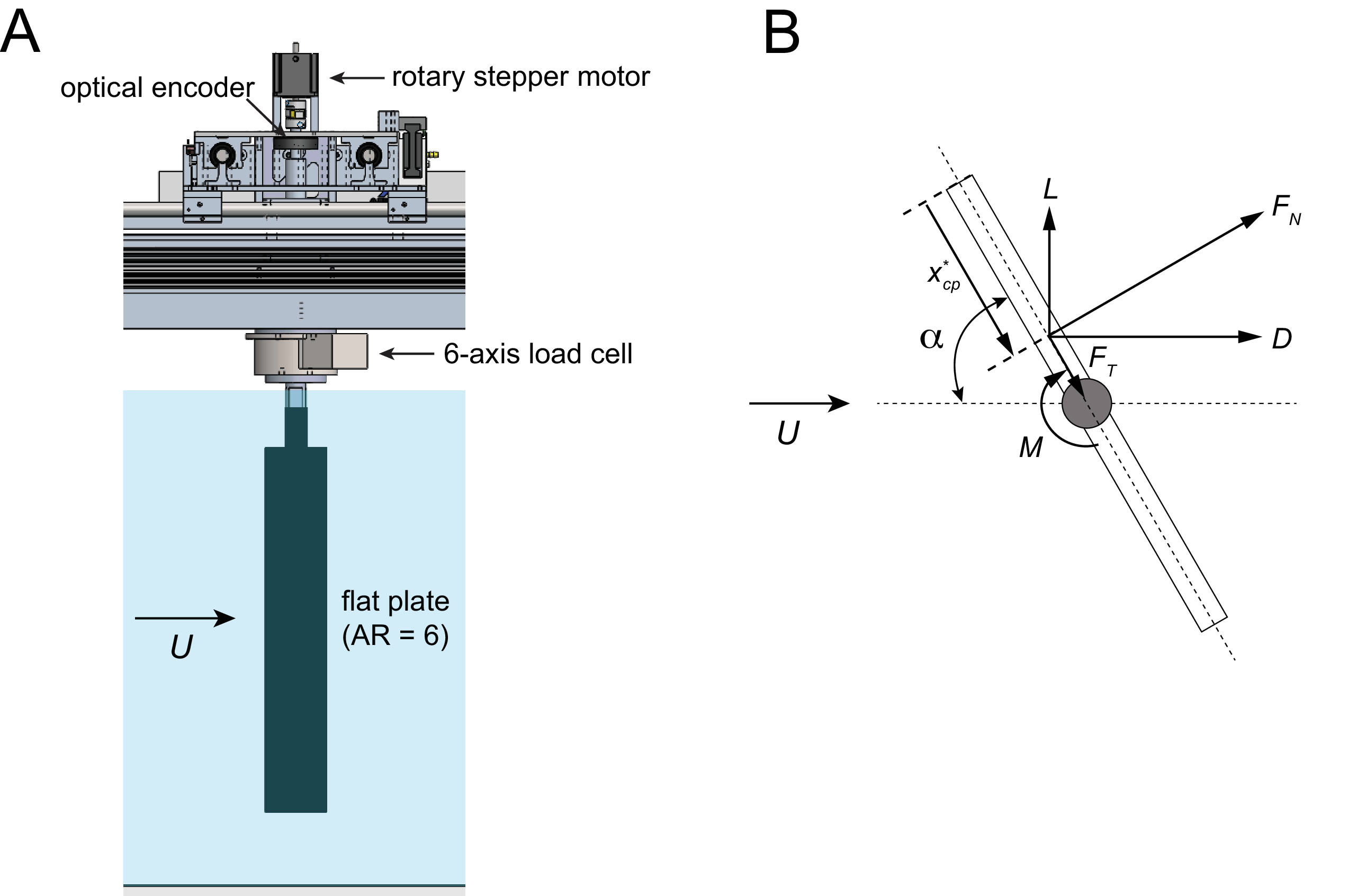}}
\caption{The experimental set-up. (A) Schematic of the experimental set-up, illustrating a flat plate (AR = 6) mounted on a 6-axis load cell about its mid-chord location. The flat plate is immersed in a uniform flow with a free-stream velocity, $U$. The present set-up allows for a direct measurement of normal and tangential forces, as well as the pitching moment exerted on a flat plate. The angle of attack, $\alpha$, is accurately measured and modulated via the optical encoder and the rotary stepper motor using an open-loop command. (B) Definition of the forces and moment imparted on a flat plate, which is subject to freestream $U$, with angle of attack $\alpha$. Lift ($L$) and drag ($D$) are approximated by decomposing the forces ($F_N$ and $F_T$) into components along the streamwise and transverse directions. The normalized center of pressure, $x^*_{cp}$, is measured from the leading edge.}
\label{fig1}
\end{figure}

All experiments were conducted in a closed-return water channel with a test section of 0.8 m in width and 0.6 m in depth. Fig.~\ref{fig1}A illustrates a flat plate, with chord, $c$ = 0.076 m, span, $h$ = 0.456 m and thickness, $\delta$ = 8 mm, rigidly affixed to the 6-axis force transducer (Delta IP65, ATI Industrial Automation), in a uniform free-stream with velocity, $U$, and density, $\rho_f$. The free-stream velocity was monitored in real time using an acoustic doppler velocimeter (Vectrino, Nortek AS). The force transducer measures the normal, $F_N$, and tangential, $F_T$, forces, as well as the pitching moment, $M$, about the rotational axis located at the mid-chord point. These forces are related to lift, $L$, and drag, $D$, forces through the geometry shown in Fig.~\ref{fig1}B, 
\begin{equation}
\begin{split}
L = F_N\cos(\alpha)-F_T\sin(\alpha), \\
D = F_N\sin(\alpha)+F_T\cos(\alpha),
\end{split}
\end{equation}
where $\alpha$ denotes the geometric angle of attack. The angle of attack is modulated by sending an open-loop command to the rotary stepper motor (Nema23 D600, Applied Motion Products) and the corresponding value is measured via an optical encoder (E3 optical encoder, US Digital). More details about the experimental setup can be found in our previous work \citep{su2019confinement,su2019resonant,su2019energy}. In this study, the Reynolds number, based on the chord length, is defined as $Re=Uc/\nu$, where $\nu$ denotes the kinematic viscosity of the fluid. Given that the chord length represents less than 10\% of the test section width, the blockage and side-wall boundary layer effects were deemed minor. Due to the use of a large aspect ratio ($AR=6$) rectangular plate, the flow associated with the present study is expected to be largely two-dimensional, and the validity of this assumption will be justified in the subsequent section. 

In the present investigation, the steady-state forces and moment associated with a steady flat plate immersed in a uniform flow were measured over a broad range of angles of attack ($\alpha$ = 0$^\circ$ $\sim$ 110$^\circ$, $\Delta\alpha=2^\circ$) at four different Reynolds numbers varying from 20,000 to 50,000 at an increment of 10,000. At each angle of attack, the measurements were recorded for 30 seconds, but only the last 20 seconds of the recording was analyzed during the post-processing in order to exclude  any transient response. The reproducibility of the results was confirmed by repeating the same set of experiments three times. 

Another useful contribution of the present study is the establishment of a simple empirical relation between the location of the center of pressure and the angle of attack. We determined the chordwise location of the center of pressure, $x^*_{cp}$ (measured from the leading edge), using the formula: 
\begin{equation}\label{eq:xcp}
    x^*_{cp} = 0.5-M/(F_Nc),
\end{equation}
where $M$ is measured pitch moment, $F_N$ is measured normal force, and $c$ is chord length. 


\subsection*{Simulation of flapping flight using proposed model}

The modeling of the flapping wing dynamical system are detailed in related papers \citep{Vejdani2019,fan2021a,fan2021b,fan2022}, but is also summarized here for completeness.

We use a quasi-steady blade element model to optimize the flight trajectory of a wing-body system subject to a prescribed flight speed.  The motions we consider are parameterized by flapping frequency, flapping amplitude, pitching amplitude. A numerical optimization is used to find wing kinematics, with lowest flapping frequency possible, that yield a stable periodic solution for a prescribed flight speed. 

Two distinct flapping wing animal - a rock pigeon and a yellow-bellied sheath-tailed bat -  are modeled to showcase the proposed quasi-steady aerodynamic models from low to high $Re$ regime. 

The morphology of the pigeon model is adapted from \citet{Tobalske1996}, with the whole mass $M=0.316$ kg, length of each wing $b/2=27.9$ cm and average wing chord $c=11.2$ cm. The morphology of the bat model is taken from the measurements of \citet{Norberg2012} with mass $M=0.068$ kg, length of each wing $b/2=30$ cm and average wing chord $c=10$ cm. Since the focus of this paper is to compare the effectiveness of the proposed high aerodynamic model $Re$, we abstracted the wings of pigeons and bats into rectangles, as the shape of the wing does not have a significant impact on the optimization results according to our previous study \cite{Vejdani2019}.

\begin{figure}[!t]   
    \centerline{ \includegraphics[trim=0 0 0 0,clip,width=6in]{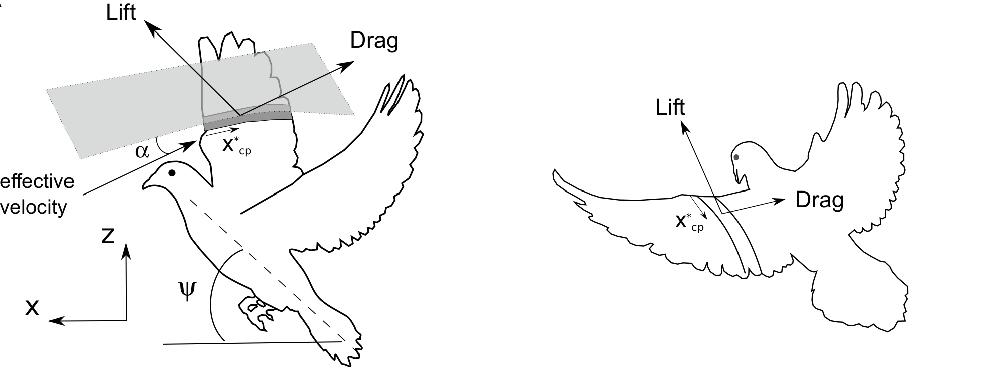}}
    \caption{Schematic of the simulated pigeon model with two translational degrees of freedom ($x-z$) and one rotational degree of freedom in pitch for the body angle ($\psi$). In addition to producing flapping and pronoation/supination motions. The simulation of bat flight shares the same segment, but has a different geometry. 
    }
    \label{fig:pigeonCartoon}
    \end{figure}

In the global inertial coordinate, \textbf{G}, the animal body has two translational degrees of freedom ($x-z$) and one rotational degree of freedom in pitch ($\psi$), where the latter accounts for the changes in the body angle, as illustrated in Fig.~\ref{fig:pigeonCartoon}. We denote the body kinematics as $\mathbf{q}_b = [x, z, \psi]^T$. The model allows modulation of its wingspan during the wingbeat cycle in order to replicate basic wing kinematics of a pigeon \citep{Vejdani2019}. For the simulation of highly-articulated bat wing, we incorporate a small amount of wing folding, and a linearly varying pitching angles from root to tip (wing twisting) \citep{fan2021b}. More specifically, the wing folding refers to additional rotation of outboard handwing with respect to inboard armwing, and is sinusoidal in time, with folding angle at mid-downstroke being zero, and mid-upstroke being maximum folding angle of $10^\circ$. The wing twisting means that the ratio of pitch angles between the wing tip and wing root is $\xi$ -- and is fixed to be 1.6 for all simulation. Note, these two wing kinematics are considered to be parameters and their values will not change during the optimization process.

Blade element theory \citep{Glauert1983,Parslew2010,fan2022} is used to model the aerodynamic forces. Specifically, the left and right wings are divided into $N = 10$ equal width segments (Fig.~\ref{fig:pigeonCartoon}),
with each segment approximating a 2D airfoil in quasi-steady state which generates a local lift and drag force vector. The total force is then a vector summation over the elemental forces on each segment in the global coordinate, \textbf{G}. It is worth mentioning that the angle of attack, $\alpha$ , seen by each wing segment is different due to the different velocities of each wing segment.
Specifically, the segment-wise lift, $\Delta{L}$, and drag, $\Delta{D}$, forces are then given as 
\begin{eqnarray}
 \Delta{L} &=& 1/2\rho C_L U_G^2 c \Delta{r}, \\
 \Delta{D} &=& 1/2\rho C_D U_G^2 c \Delta{r},
\end{eqnarray}
where $\rho$, $c$, $\Delta{r}$ and $U_G$ are the air density, chord length, segment width, and airspeed at each segment in the global coordinate system, \textbf{G}. The lift and drag on each segment are located at the quarter-chord. Note, the aerodynamic coefficients ($C_L$ and $C_D$) used in the lift and drag model are taken from \citet{Dickinson1999} and current study, respectively.

To construct the equations of motion, we assume the left and right wing motions to be symmetric. Thus, the dynamical model has 5 degrees of freedom in total: the animal position, $x, z$, body pitch angle $\Psi$, and the wing root flapping and pitching amplitudes, $\phi_r$ and $\theta_r$, respectively.

Along with their derivatives, they form a 10 by 1 state vector $\mathbf{q}$ with which we construct the Lagrangian equation of motion:
\begin{eqnarray}
\mathbf{D}(\mathbf{q}) \ddot{\mathbf{q}} + \mathbf{C}(\mathbf{q},\dot{\mathbf{q}}) \dot{\mathbf{q}} + \mathbf{g}(\mathbf{q}) = \mathbf{\tau} + \mathbf{f}_{aero}, 
\label{eqn:lagEoM}
\end{eqnarray} 
where $\mathbf{D}$, $\mathbf{C}(\mathbf{q}$, $\dot{\mathbf{q}})$ and $\mathbf{g}(\mathbf{q})$ are the mass matrix, centrifugal matrix and gravitational vector, respectively. The generalized torque, $\mathbf{\tau} = [ 0_{1\times3},\tau_{w}]^T $, denotes the internal forces between the wing and the body, while $\mathbf{f}_{aero} $ represents the external aerodynamic forces and the ensuing torques felt by both left and right wings.

Mathematically, steady forward flight is a solution to the equations of motion that satisfy 
\begin{eqnarray}
[ \mathbf{q},\dot {\mathbf{q}} ]_t = [ \mathbf{q}, \dot {\mathbf{q}}]_{t+T} + \Delta\cdot{T},
\label{eqn:POrbit}
\end{eqnarray}
where $t$ is an arbitrary instant of time, $T$ is wingbeat period.  $\Delta$ is a 10-valued vector, $\Delta = [U_x,U_z,0,0,0,0,0,0,0,0]^T$. Here, $U_x$ and $U_z$ are the cycle-averaged velocity components along the forward ($X$) and vertical ($Z$) directions in the global coordinate system, \textbf{G}, and are defined as inputs. If $U_x, U_z = 0$, this becomes the more familiar expression for hovering flight. 

In this paper, we focus on forward and level flight  ($U_z = 0$). As pointed out earlier, there are many sets of possible wing kinematic parameters that may satisfy the constraints (Eqn~\ref{eqn:POrbit}), given appropriate initial conditions ($\mathbf{q}_0,\dot{\mathbf{q}}_0$). Therefore, an optimization scheme is employed that searches over all (realistic) candidates of wing kinematics that result in the prescribed forward speed, $U_x$, and finds the one with minimum frequency, $f$. A typical periodic cycle for forward level flight is given in Fig.~\ref{fig:periodicOrbit}.

At each time $t$, the instantaneous total power, $P_{tot}$, can be obtained as \citep{Parslew2015,fan2022}
\begin{equation}\label{eqn:pwrCalc}
 P_{tot}(t) = \boldsymbol{\tau}_{w}\dot{\mathbf{q}}_{w},
\end{equation}
where $\dot{\mathbf{q}}_{w}$ are the generalized velocities for the two wings and $\boldsymbol{\tau}_{w}$ are the generalized internal forces of the wing-body system.

\section*{Results}

\subsection*{Quasi-steady aerodynamic model at moderate Reynolds numbers}

\begin{figure}[!t]    
\centerline{ \includegraphics[trim=0 0 0 0,clip,width=6.5in]{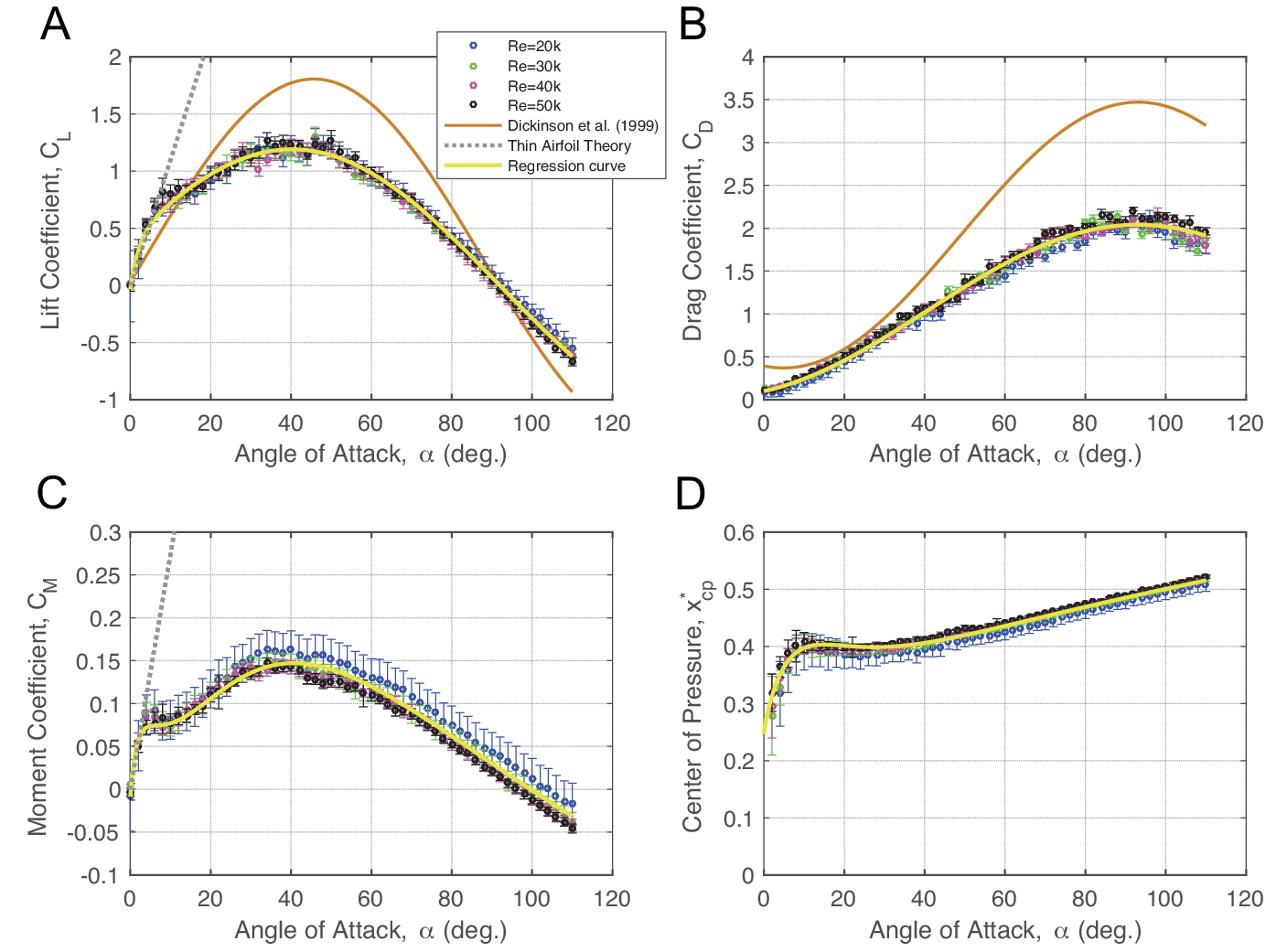}}
\caption{The evolution of normalized lift (A), drag (B), moment (C), and center of pressure (D) as a function of the angle of attack, $\alpha$, at four distinct Reynolds numbers. Steady-state forces and moment are appropriately normalized by the inertial forcing of the fluid flow. The error bar associated with each data point represents the standard deviation of the time-averaged measurement. Thin airfoil theory predictions of ${C}_{L}$ (i.e., $2\pi\alpha$) and ${C}_{M}$ (i.e., $\frac{\pi}{2}\alpha$) are plotted with the gray dotted lines in (A) and (C), respectively. The  regression curves for $C_L$, $C_D$ and $x^*_{cp}$ are plotted with the solid yellow curves, where the corresponding equations are provided in Eqns \ref{eq:forces}, \ref{eq:forces2} and \ref{eq:CoP}. The regression curve for $C_M$ is deduced analytically using the expressions of $C_L$, $C_D$ and $x^*_{cp}$ (discussed further in the text).}
\label{figClCdCmCop}
\end{figure}

The lift and drag coefficients as a function of the angle of attack are presented in Fig. \ref{figClCdCmCop}A and B, respectively, where the lift and drag coefficients are defined as

\begin{eqnarray}\label{eq:force def}
    C_L=2L/\rho_f U^2ch, \\
    C_D=2D/\rho_f U^2ch.
\end{eqnarray}

The lift and drag coefficients are well approximated by the following empirical relationships in terms of the angle of attack, $\alpha$ (in degrees):

\begin{eqnarray}
{C}_{L}(\alpha)&=&-0.44e^{-0.47\alpha}+1.19\sin(1.74\alpha+20^{\circ}) \label{eq:forces},\\
{C}_{D}(\alpha)&=&1.04+\sin(1.72\alpha-70^{\circ}),  \label{eq:forces2}
\end{eqnarray}
where, in each case, the best-fit regression curve with a 95\% confidence interval yielded an R-squared value greater than 0.995. 
As illustrated in Fig.~\ref{figClCdCmCop}A, the $C_L$ curve follows closely the steady thin airfoil theory prediction with a slope of d$C_L$/d$\alpha$ = $2\pi$ at small angles of attack ($\alpha<8^{\circ}$). Upon the attainment of maximum lift coefficient at approximately $\alpha \approx 40^{\circ}$, the lift curve decays monotonically as $\alpha$ is further increased. The low-Reynolds number quasi-steady model of \citet{Dickinson1999} is shown as an orange line for comparison.

In Fig. \ref{figClCdCmCop}B, the drag coefficient increases monotonically from approximately 0.1 at $\alpha$ = 0 to a maximum value of 2.0 at $\alpha \approx 90^{\circ}$. Although it has a similar trajectory, it lies below the low-Reynolds number model (orange curve) for all values of angle of attack.

The pitch moment coefficient, $C_M$, about the mid chord,
\begin{equation}
C_M=\frac{M}{0.5 \rho_f U^2c^2h} , 
\end{equation}
is shown in Fig.~\ref{figClCdCmCop}C. It is observed that the pitch moment data follow a very similar trend to the lift coefficient. In particular, the $C_M$ curves show a slope of d$C_M$/d$\alpha$ = $\pi/2$ at small angles of attack, closely following the steady thin airfoil theory prediction.

Finally, the evolution of center of pressure, $x^*_{cp}$ (Eqn~\ref{eq:CoP}),  as a function of $\alpha$ is illustrated in Fig.~\ref{figClCdCmCop}D at four distinct Reynolds numbers. We can see that all four cases collapse well. An empirical fit of center of pressure, $x^*_{cp}$, is given by  
\begin{equation}\label{eq:CoP}
\begin{split}
{x}^*_{cp}(\alpha)=0.247+0.016\alpha^{0.6}+0.026{\alpha}e^{-0.11\alpha},
\end{split}
\end{equation}
which is the yellow solid curve in Fig.~\ref{figClCdCmCop}D. One can evidently infer that the current model predicts that $x^*_{cp}$ approaches the aerodynamic center at 0.25 (quarter chord) in the limit as $\alpha$ tends to zero, which agrees well with the linear inviscid theories \citep[e.g.,][]{Wagner, Theodorsen, Garrick}. With the knowledge of center of pressure, the moment coefficient, $C_M$, about any arbitrary point can be easily approximated using Eqns \ref{eq:forces}, \ref{eq:CoP} and
\begin{equation}
    C_M = C_N(x^* - x^*_{cp}),
\end{equation}
where $C_N$ is the normal force coefficient, defined as $C_N$ = $C_L\cos(\alpha)$ + $C_D\sin(\alpha)$, which is the measured normal force coefficient of $F_N$, and $x^*$ denotes the arbitrary location of the moment axis with respect to the leading edge (e.g., $x^*$ = 0.25 for the quarter-chord point). Using this formulation, we evaluated the moment coefficient about the mid-chord point ($x^*$ = 0.5), and the result is plotted in Fig.~\ref{figClCdCmCop}C with a yellow curve, revealing an excellent agreement with the experimental measurements (plotted with open circles).

\subsection*{Application in modeling bird and bat flight}

\begin{figure}
    \centering
    \includegraphics[width=6in]{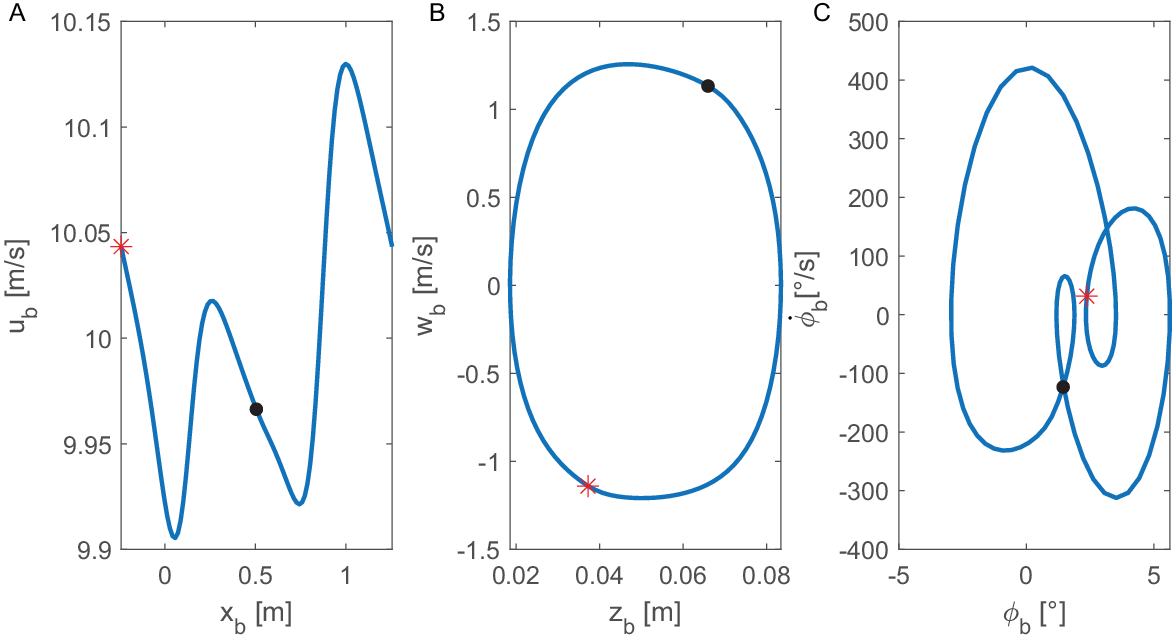}
    \caption{A sample periodic orbit of the flapping wing dynamical system. The motion of the wing-body systems - vertical displacement and body pitching angle are cyclic, but there is net accumulation in the forward direction. (A) Horizontal velocity, $u_b$, and dispacement, $x_b$; (B) vertical velocity, $w_b$, and displacement, $z_b$; (C) body pitch velocity $\phi_b$, and pitch angle, $\phi_b$. Asterisk represents beginning of downstroke, and black solid circle represent beginning of upstroke. The average flight speed $U = 10$m/s.}
    \label{fig:periodicOrbit}
\end{figure}

Using the aerodynamic models from \citet{Dickinson1999} and current study (Eqn \ref{eq:forces} and \ref{eq:forces2}), we apply our dynamic model (Eqn \ref{eqn:lagEoM}) to simulate the forward flight of an idealized  pigeon and bat, as described earlier. Given a flight speed, the computational model finds the minimum possible flapping frequency among all candidate wing kinematics; a sample solution is illustrated using phase plots in Fig.~\ref{fig:periodicOrbit}. The forward ($u_b$), vertical ($w_b$) velocities and body pitching rate ($\dot{\phi}_b$) vary within a cycle, and because the target speed is level and straight, the phase plots of $w_b$-$z_b$ and $\dot{\phi}_b$-$\phi_b$ form closed loops. In contrast, the phase plot of $u_b$-$x_b$ does not form a closed loop, reflecting the forward translation.

\begin{figure}
    \centering
    \includegraphics[width=6in]{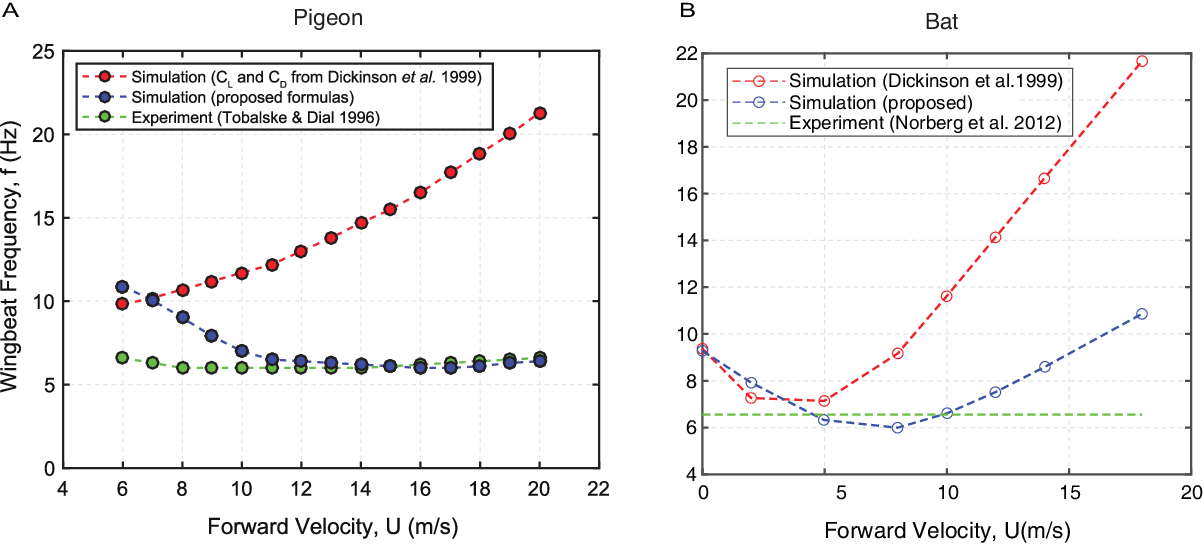}
    \caption{Predicted flapping frequency of (A) a pigeon and (B) a bat as a function of flight speed. The dashed green line is experimentally measured flapping frequency. Red dots represent the  wingbeat frequencies evaluated using the quasi-steady aerodynamic formulations of \citet{Dickinson1999}. Blue dots designate the predicted wingbeat frequencies using the proposed aerodynamic formulas. Only the aerodynamic models are different between the two simulation.  }
    \label{fig:freqVSSpeed}
\end{figure}

We then vary the prescribed flight speed to study how the aerodynamic models impact the optimized flapping frequency over a range of physically plausible speeds (Fig.~\ref{fig:freqVSSpeed}). For the flight of a pigeon in Fig.~\ref{fig:freqVSSpeed}A, at high forward velocities ($U >$ 10 m/s), the aerodynamic model in current study (blue dots) perform remarkably well in predicting the measured wingbeat frequencies (green dots) by \citet{Tobalske1996}, while the formulations of \citet{Dickinson1999} (red dots) result in a poor agreement with the empirical observations. For the flight of a bat (Fig.~\ref{fig:freqVSSpeed}B), at low forward speeds ($U < 5$ m/s), the two models predict similar flapping frequency, but as the flight speed picks up ($U > 5$ m/s), the difference in frequency between the two models becomes much more pronounced. Live animals typically fly with a constant flapping frequency around $6$~Hz \citep{Norberg2012} and is invariant of flight speed.  The predicted flapping frequency given by our proposed model agree very well with recorded observation data, especially around $U = 5 - 10$ m/s.

\begin{figure}
    \centering
    \includegraphics[width=6in]{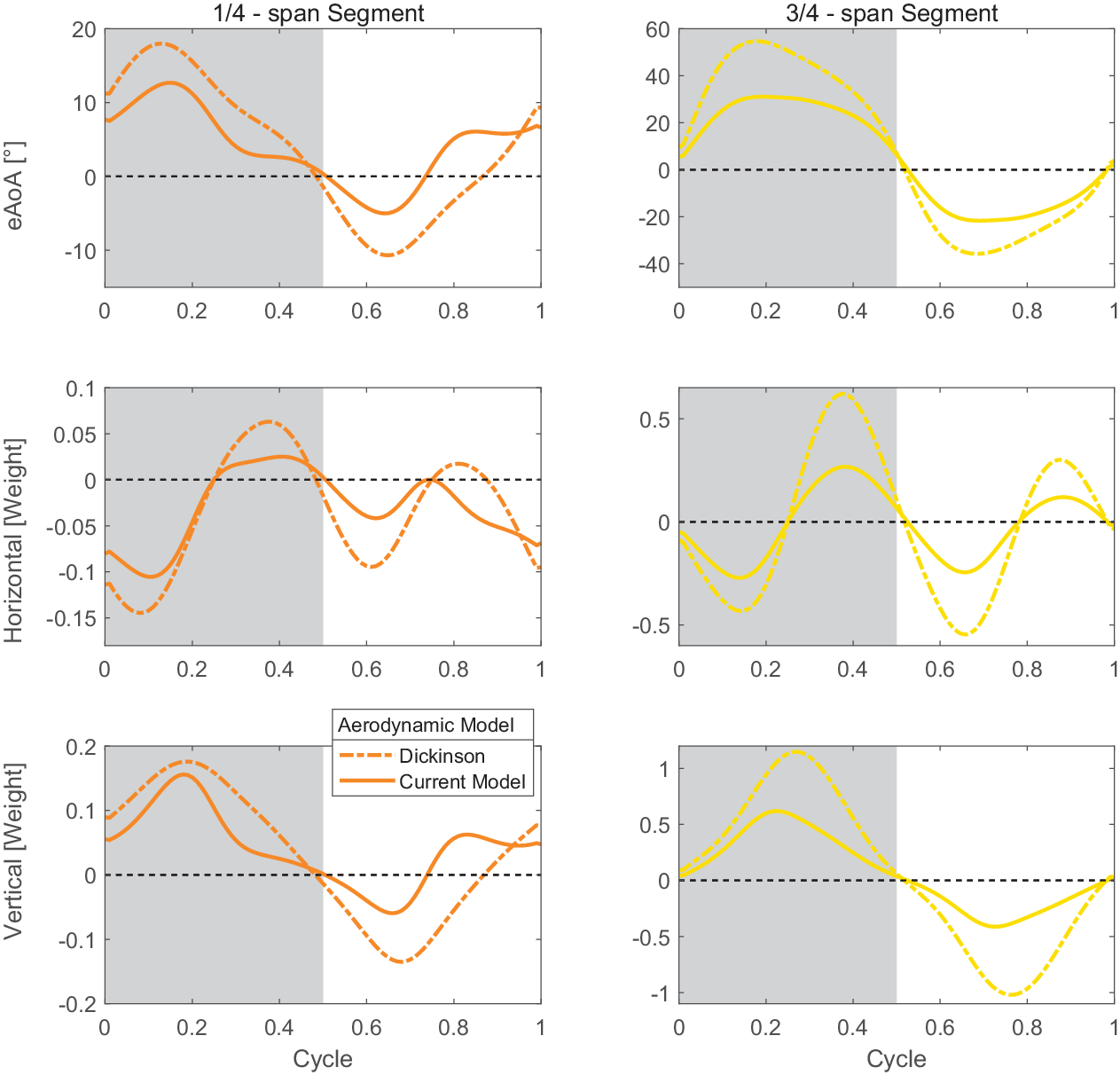}
\caption{Effective angles of attack, horizontal and vertical forces produced by $1/4$ and $3/4$-span wing segments. The freestream $U = 10$ m/s, and the dashed and solid lines are time series using aerodynamic models proposed by \citet{Dickinson1999} and the current study (Eqn.~\ref{eq:forces}), respectively. The shaded area represents downstroke. The forces are normalized by the weight of the animal, and time is normalized by one flapping period.}
    \label{fig:aoaForces}
\end{figure}

Fundamentally, the predicted differences in flapping frequency are manifestation of different aerodynamic forces experienced by the wings. We choose a representative case of $U = 10$ m/s for the bat flight in Fig.~\ref{fig:aoaForces} to illustrate the evolution of forces due to the different aerodynamic models. The new aerodynamic model (Eqn~\ref{eq:forces} and \ref{eq:forces2}) predicts smaller effective angles of attack for both inboard ($1/4$-span) and outboard ($3/4$-span) of the wing, and as a result, smaller horizontal and vertical forces are observed.

\begin{figure}
    \centering
    \includegraphics[width=3in]{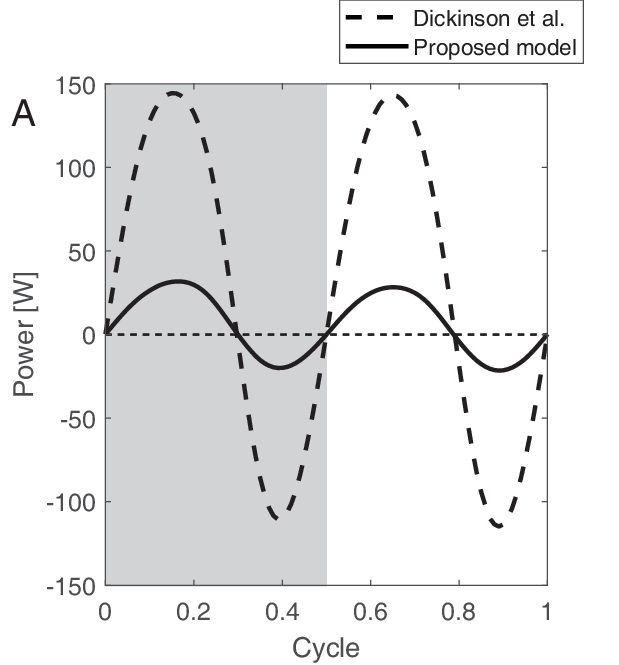}
    \caption{Predicted total power for one cycle at forward flight speed of $U = 10$ m/s. Dashed and solid lines represent results using models from \citet{Dickinson1999} and current study, respectively. Shaded area represent downstroke. Time is normalized by one flapping period.}
    \label{fig:pwrCmp}
\end{figure}

Another important distinction between the two aerodynamic models concerns flight energetics. As seen in Fig.~\ref{fig:pwrCmp}, the over-prediction in flapping frequency (Dickinson and Sane model, $U = 10$ m/s in Fig.~\ref{fig:freqVSSpeed}) results in an unrealistic maximum power of $150$ W, whereas the proposed model predicts a peak power of only $30$ W.

\begin{figure}
    \centering
    \includegraphics[width=3in]{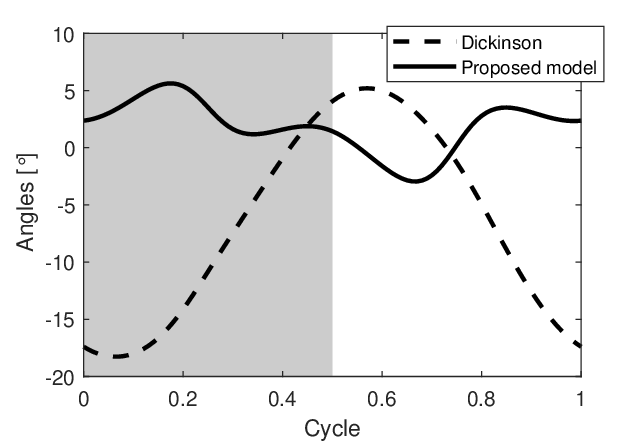}
    \caption{Predicted body pitching angle variation within one cycle. Shaded area represent downstroke.}
    \label{fig:bdPtchCmp}
\end{figure}

Finally, one important aspect that relates to applications in flight robotics is the body pitching angle variation within a cycle - which may impact onboard camera measurement, inertia measurement unit (IMU) sensoring, thus detrimentally impacting the state estimation. In Fig.~\ref{fig:bdPtchCmp}, we see the range of pitching angles using Dickinson and Sane model can be as large as $25$ degrees, whereas the current model bounds the pitching variation within $10$ degrees.
  
\section*{Discussion}

\subsection*{Aerodynamic forces at moderate Reynolds number}

In the current study, the aerodynamic forces of a rectangular foil are characterized at a series of angles of attack over a range of Reynolds numbers (Re = 20,000 to 50,000). The magnitudes of lift coefficient, $C_L$, and drag coefficient, $C_D$, are found to be comparable to those obtained from steady-state force measurements of a pitching flat plate (Re = 20,000) \citep{Granlund2013} and from measurements on a translating two-dimensional rectangular wing at various fixed angles of attack (Re $\sim$ 200) \citep{Dickinson1993}. The peak of lift coefficient, $C_L$ in Fig.~\ref{figClCdCmCop}A, is achieved at around $40^{\circ}$ in the current study, which is also in line with experimental studies using flat plates \citep{Dickinson1993,usherwood2009aerodynamic,Granlund2013}, insect wing models \citep{Dickinson1999}, and a dried pigeon wing \citep{usherwood2009aerodynamic}. 

The maximum drag coefficient, $C_D$ in Fig.~\ref{figClCdCmCop}B, conforms very well with the theoretical prediction of 1.95, wherein the drag coefficient associated with a two-dimensional flat plate facing perpendicular to the incoming flow can be approximated as $C_D = 1.95+50/Re$ \citep{Ellington1991}. The rough shape of the drag coefficient ($C_D$) in the current study is well approximated with a simple harmonic relationship, and resembles the functional form derived in the experimental work of \citet{Dickinson1999}, where steady-state translational force measurements (i.e., $C_L$ and $C_D$) were conducted on a dynamically scaled-up $\it{Drosophila}$ wing at a Reynolds number of approximately 140, much smaller than that in the current study (Re = 20,000 to 50,000).

However, the general characteristics of the present force profiles differ conspicuously from those of \citet{Dickinson1999}. Firstly, the slope of our $C_L$ values conform with the thin airfoil theory predictions for small angles, \citet{Dickinson1999} largely underestimates the theoretical prediction of d$C_L$/d$\alpha$ = $2\pi$ with a value of approximately 3.8. This discrepancy is likely due to the experiments in \citet{Dickinson1999} surveyed a low $Re$ of $\mathcal{O}(10^2)$, which does not satisfy the inviscid flow assumption --- the basis of the thin airfoil theory \citep{Bisplinghoff}. Secondly, the maximum $C_L$ and $C_D$ obtained in the present study are approximately 45\% and 65\% smaller, respectively. Higher steady-state aerodynamic forces associated with a rotating 3D wing is due, in part, to the formation, growth and subsequent stabilization of an attached leading-edge vortex (LEV) on the upper surface (or suction/dorsal side, used synonymously) of the wing, as previously noted by a number of investigators \citep[e.g.,][]{Maxworthy1981, ellington1996leading, Dickinson1999}. The general consensus in the archival literature is that the parameter of rotation (e.g., revolving, flapping and pitching) plays an important role in the formation and subsequent stabilization of the LEV \citep[e.g.,][]{Lentink2009a, Wong2015, Onoue2016}, and that this enhanced vortex stability is directly correlated to the generation of strong aerodynamic forces \citep{Granlund2013, Harbig2013,ribeiro2021wake,otomo2021unsteady} and pitching moments \citep{Onoue2015, Onoue2016,zhu2020nonlinear} that far exceed their static counterparts, easily by a factor of 1.5 to 2.5 in magnitude.

Indeed, the fundamentals of vortex dynamics holds the key to understand the differences of the maxima presented by the two studies. \citet{Gharib1998} used a piston-cylinder apparatus to study vortex formation, and quantitaively recorded that the vortex core will ``saturate'' in strength, and reject additional vorticity from the shear layer, after a nondimensional timescale, coined as the formation time $\hat{T} = UT/D$, exceeds a critical value, where $U$ is the freestream velocity, $T$ the time it takes for a complete vortex ring to shed, and $D$ the characteristic length. For the cylinder apparatus, the vortex sheds or ``pinches off'' after $\hat{T} = 4$. Some constant correction may be needed for other geometries than the cylinder \citep{Dabiri2009}. Both Dickinson and the present study observe effect of the LEV, but centripetal force is present in the former study due to flapping of wings, thus a spanwise flow is present which is absent in our 2D case. Indeed, \citet{ellington1996leading} first observed the spanwise flow in flapping wing flight of $Re \sim 10^3$, which would transport additional vorticity outboard before finally being convected to wake as part of tip vortex, thus stabilizing the LEV (no satuation of vorticity) on the dorsal side during downstroke. \citet{Lentink2009} then demonstrated that this stabilizing mechanism for LEV holds for a much wider range of $Re \sim 10^2 - 10^5$ -- touching on the lower $Re$ range of this study. Later, \citet{Harbig2013} found that high aspect ratio wing (the extreme is a 2D wing like ours) features reduced LEV circulation, and thus less lift coefficients. In our study, the wing is of high aspect ratio, and has an end plate to reduce tip effects, and thus the experiment is more close to a two-dimensional  flow (infinite AR) without spanwise flow, so the LEV is likely become unstable and shed much earlier than \citet{Dickinson1999}, where their experiment features a rotating flapping wing with finite AR at $Re = 10^2$, where spanwise flow convects extra vorticity outboard to tip vortex, and thus delay the shedding of LEV, which contributes to higher peak $C_L$ and $C_D$.

It should be noted that the forces measured in the present study are ``quasi-steady" such that the concomitant shedding of von K\'{a}rm\'{a}n vortices behind the plate does not induce any measurable time-dependency in the forces and the pitching moment imparted on the wing. This is presumably because the formation of these large-scale vortices occurs in the far-wake region approximately 1.5 to 2 chord lengths downstream of the plate (for Reynolds numbers on the order of $\mathcal{O}(10^4)\sim\mathcal{O}(10^5)$), as discussed in \citet{Onoue2016} and \citet{Liu2017}. The validity of the present assertion is also supported by small standard deviations associated with the force and moment data presented in Fig.~\ref{figClCdCmCop}, reflecting minor unsteadiness in the measurements. 

This study also contributes a simple empirical relationship between the center of pressure's location and the angle of attack. \citet{han2015} was one of the early studies to demonstrate the importance to consider variation in center of pressure using quasi-steady modeling. By simulation of a hawkmoth scale hovering flight ($Re \sim 10^2$), they observed a more consistent periodic flight characteristics when the center of pressure of the insects' wing is allowed to be a function of angle of attack. In a much higher $Re \sim 10^5$, when the wing inertial was ommited \citep{Windes2018}, the simulated body trajectories of a forward flight bat achieved only partial success, subsequently \citet{Fan2021} demonstrated how the pitch moment of inertia of the wing-body system is critical to achieving high-fidelity modeling of the dynamic system. The formulation provided here may thus serve as an important link for high $Re$ case.

Note that for both  \citet{han2015} ($Re \sim 10^2$) and present study ($Re \sim 10^5$) (Fig.~\ref{figClCdCmCop}D), the center of pressure starts around quarter-chord at small angle of attack $\alpha$, which agrees well with thin airfoil theory \citep{Bisplinghoff}, and monotonically shifts backwards towards trailing edge, and stays around $x/c = 0.4$ until $\alpha$ reaches $\sim 40^\circ$. In the meantime, the moment coefficient $C_M$ keeps increasing with $\alpha$, but also peak until $\sim 40^\circ$, which suggests that the influence of LEV on force production decreases past $\alpha \sim 40^\circ$.

\subsection*{Animal flight simulations and energetics}

Equipped with the proposed model, which is suitable for moderate to high $Re$, the resulting flapping frequency of both the pigeons and the bats in the simulations are closer to experimental measured values (Fig.~\ref{fig:freqVSSpeed}), especially at high velocities or Reynolds number. The fixed flapping frequency with changing flight speed is often observed in live animal flight, as animals modulate their flight speed by more localized and precise control of their wings such as armwing camber control, handwing twisting and folding, etc \citep{fan2022}.

The average lift over a cycle has to equal the weight of the animal, and thrust must balance drag to achieve forward and level flight. Using the proposed aerodynamic model in the current study, extra lift coefficient is available at small angle of attack $\alpha = 0 \sim 13 ^\circ$ (Fig.~\ref{figClCdCmCop}A) with small penalty of drag (Fig.~\ref{figClCdCmCop}B) -- in other words, our proposed model features a high $C_L/C_D$ regime for small $\alpha$. As a result, our algorithm found a solution where the armwing requires minimal pitching to leverage these advantages, which manifests as a smaller effective angle of attack throughout the cycle (solid line in top left panel of Fig.~\ref{fig:aoaForces}). In a loose language, the armwing segments feature a small nose-up pitch throughout the cycle, which generates lift with small drag penalty. On the other hand, when the Dickinson and Sane low Reynolds model \citep{Dickinson1999} is applied at this high Reynolds scenario (pigeon or bat flight), both $C_L$ and $C_d$ are small within this range of angle of attack ($\alpha = 0 \sim 13^\circ$). Consequently, the algorithm can only rely on a larger $\alpha$ (dashed line in top left panel of Fig.~\ref{fig:aoaForces}) to generate the required lift, resulting in a higher drag penalty. The outboard wing segment (3/4-span) is further away from wing root, and thus has a larger $\alpha$ compared to inboard wing segment (1/4-span), and would not utilize the small $\alpha$ (or high $C_L/C_D$ value) ``perk'', which is true for either model. 

Interestingly, \citet{shyy2013} pointed out that armwing (represented as 1/4-span wing segment) is mainly responsible for lift generation, and handwing (3/4-span wing segment) for thrust. This statement is supported by our simulation using both aerodynamic models. In Fig.~\ref{fig:aoaForces}, we indeed see that the armwing generates a net positive lift compared to the handwing, and that for horizontal force, the armwing features more drag.

The significance of the current proposed model is also evident in the power consumption. Both CFD \citep{Windes2018} and quasi-steady modeling \citep{fan2022} of similar sized bats show instantaneous power consumption $<4$ W, which is lower than that predicted by either of the two aerodynamic models (Fig.~\ref{fig:pwrCmp}). Our proposed model nevertheless provides a much closer power prediction compared to Dickinson and Sane model. Note that around $U = 10$ m/s, their model predicted almost twice as much flapping frequency in Fig.~\ref{fig:freqVSSpeed}, and because power roughly scales quadratically with flapping frequency \citep{fan2022}, the instantaneous power is thus four times larger (Fig.~\ref{fig:pwrCmp}).

It is noteworthy that bats typically have a very small body pitching variation in flight, only by a few degrees \citep{Fan2021}, which is accurately reflected using our proposed model. This stability is primarily due to the extra lift with small drag penalty at small angle of attack, allowing the wing to generate sufficient cycle-averaged lift and maintain thrust-drag balance. As a result, the body does not need to oscillate as violently so the wing may reach large angle of attack (top left and right panels in Fig.~\ref{fig:aoaForces}). 

Finally, it is important to highlight that the differentiable nature of the proposed formulas makes them highly suitable to be used in dynamical simulations that utilize gradient-based optimization techniques to determine, for example, power-optimal kinematics of avian wings for different modes of flight. 


\section*{Acknowledgements}
The authors would like to thank the members of the Energy Harvesting group at Brown University for their contributions to the development of the experimental apparatus.

\section*{Competing interests}
The authors declare no competing financial interests.

\section*{Author contributions}
Y.S., X.F, K.O and H.V. drafted the paper; K.O., H.V. Y.S. designed and executed the experiments; K.O. characterized the aerodynamic models;  X.F. and H.V. performed numerical simulations, K.B. supervised and managed the project, reviewed and edited the draft, and acquired financial support on the project.  

\section*{Funding}
The original experiments were supported by the Department of Energy's Advanced Research Projects Agency-Energy (ARPA-E), Grant No. DE-AR0000318.  Subsequent data analysis and manuscript preparation were supported by NSF Grant IOS-1930924 and ONR Grant N00014-21-1-2816.

\bibliographystyle{apalike}
\bibliography{jfm-instructions}

\end{document}